\documentclass[floatfix,%
 reprint,
nofootinbib,
 amsmath,amssymb,
 aps,
]{revtex4-2}

\usepackage{graphicx}
\usepackage{braket}
\usepackage[T1]{fontenc}
\usepackage[utf8]{inputenc}
\usepackage{footnote}
    \usepackage{framed}
\usepackage{dcolumn}
\usepackage{bm}
\usepackage{graphicx}
\usepackage{subcaption} 
\usepackage{float}
\usepackage{microtype}
\usepackage{mathrsfs}
\usepackage[colorlinks,citecolor=blue,urlcolor=blue,linkcolor=blue]{hyperref}

\begin{document}

\preprint{APS/123-QED}

\title{Future Rip Scenarios in Fractional Holographic Dark Energy}

\author
{Ayush Bidlan$^{1}$}
\email{i21ph018@phy.svnit.ac.in}
\author{Paulo Moniz$^{2}$}
\email{pmoniz@ubi.pt}
\author{Oem Trivedi$^{3}$}
\email{oem.trivedi@vanderbilt.edu}
\affiliation{%
 $^{1}$Department of Physics, Sardar Vallabhbhai National Institute of Technology, Surat 395007, Gujarat, India,
}%
\affiliation{$^{2}$Departamento de Física, Centro de Matemática e Aplicações (CMA-UBI), Universidade da Beira Interior, Marquês d’Avila e Bolama, 6200-001 Covilhã, Portugal,}
\affiliation{$^{3}$Department of Physics and Astronomy, Vanderbilt University, Nashville, TN, 37235, USA}%

\date{\today}

\begin{abstract}
In this paper, we investigate the occurrence of late-time cosmological singularities, namely, the rip scenarios within the framework of interacting Fractional Holographic Dark Energy (FHDE). We start our investigation of rip scenarios with the Granda-Oliveros (GO) cutoff, i.e., $L=(\gamma H^{2}+\delta\dot{H})^{-\frac{1}{2}}$, and highlight the range of allowed $\alpha$ (Lévy's index) values for which big, little and pseudo rip can occur. In particular, we highlight the occurrence of a big rip for fractional values of the Lévy's index in the allowed range $1<\alpha\leq2$. Moreover, within a similar cosmological setting, we conclude that the occurrence of a pseudo-rip requires Lévy's index to be $\alpha>2$. Therefore, we reject the possibility of pseudo-rip within the GO cutoff. Furthermore, we also demonstrate that the occurrence of the little rip in FHDE equipped with a GO cutoff is rather contrived and requires a specific functional form of the IR cutoff $L\sim(\gamma H^{2}+g(H))^{-\frac{1}{2}}$, which belongs to a larger class of Nojiri-Odintsov (NO) cutoffs. To extend our perspective beyond the GO cutoff, we investigate the interacting FHDE framework equipped with the Hubble cutoff, i.e., $L=H^{-1}$, in developing an ansatz-based approach to the little and pseudo-rip singularities as they fail to appear in the GO cutoff. Within this approach, we invoke the expression of the Hubble parameter, $H(t)$, which corresponds to the little and pseudo-rip, into the cosmological parameters such as the Equation of State (EoS) and Squared Sound Speed (SSS) as a function of cosmic time $t$. Specifically, we produce numerical plots of these parameters in both linear and non-linear $Q$ regimes, which supplement our theoretical findings. In summary, our results highlight the occurrence of little and pseudo-rip singularities within a Hubble cutoff for a non-linear $Q$ term within the FHDE framework.
\end{abstract}


\maketitle

\section{Introduction}\label{Section 1}
Observations in modern cosmology appears to suggest the presence of an evolving dark energy responsible for the varying acceleration at which the Universe is growing (expansion) \cite{SupernovaSearchTeam:1998fmf, SupernovaCosmologyProject:1998vns, Sousa-Neto:2025gpj, DESI:2024mwx, Abdul_Karim_2025, Li:2024qso, Gialamas_2025,chaudhary2025evidenceevolvingdarkenergy}. The traditional method for addressing late-time cosmic acceleration is based on the addition of a cosmological constant, denoted by $\Lambda$, to the field equations of the General Theory of Relativity \cite{Einstein:1917ce, Weinberg:1988cp}. This constant serves as a type of dark energy that permeates space, inducing peculiar dynamics that oppose Newtonian attractive gravity, thereby fueling an accelerated expansion of the Universe. This scenario constitutes our current understanding of the Universe on large scales and is known as the $\Lambda$CDM (Cold Dark Matter) model. It is pertinent to emphasise that the cosmological constant $\Lambda$ represents the simplest theoretical framework in order to explain the observational evidence of an accelerating Universe. However, very recent observations from the Dark Energy Spectroscopic Instrument Data Release 2 (DESI DR2) suggest a considerable preference for time-dependent dark energy models over $\Lambda$CDM \cite{cortês2025desisdr2exclusionlambdacdm, Abdul_Karim_2025, Efstathiou:2024dvn}. In particular, the highest quoted $4.2\sigma$ comes from DESI + Planck + DESY5 supernovae, but conservative reanalyses suggest $\sim3\sigma$ as a robust lower bound due to methodological variations. No single combination exceeds $5\sigma$, so the $\Lambda$CDM model remains viable but increasingly challenged \cite{cortês2025desisdr2exclusionlambdacdm, Efstathiou:2024dvn}. Such a direct observational discovery opens up a whole new realm of possibilities for the right framework of dark energy \cite{Brax:2025ahm,Brandenberger:2025hof, Brax:1999gp, Trivedi_2024, Bidlan_2025, Brevik:2024ozg}. Moreover, dark energy is characterised by negative pressure responsible for accelerated expansion of the Universe. In terms of the standard energy conditions in general relativity \cite{Hawking:1973uf}, dark energy must violate the strong energy condition $\rho+3p>0$, $\rho>0$. Assuming a barotropic Equation of State (EoS) of matter in the Universe, $p=w\rho$ with a constant EoS $w$, where $p$ and $\rho$ are the pressure and the density of dark energy, respectively, it requires that $w<-1/3$. However, according to more recent observations \cite{SDSS:2003eyi}, dark energy is even \textit{more} biased towards larger negative values of the barotropic index $w\lesssim-1$. This directly implies that it would lead to the violation of the null energy condition $\rho+p>0$ and consequently all other energy conditions as well. Dark energy of this type was termed as ``phantom" \cite{Caldwell_2003, Nojiri_2003, Hannestad_2002}. Notably, phantom dark energy models of the Universe admit various types of rip singularities \cite{Caldwell_2003, Nojiri_2003}.

\bigskip

Within the scope of this work, we find it promising to explore an interdisciplinary approach to dark energy, which we believe could be fruitful. Specifically, our main motivation towards such an approach is to explore the interplay of non-local features induced by fractional integro-differential operators in the HDE framework \cite{Trivedi_2024} to investigate the occurrence or avoidance of late-time rip singularities such as the big rip, little rip and the pseudo-rip. Let us first highlight that we particularly combine the features of Fractional Calculus (FC) \citep{ortigueira2011fractional,ortigueira2017fractional,ortigueira2012relation,ortigueira2023fractional,valerio2023variable,bengochea2023operational,ortigueira2024factory}  and Holographic Principle (HP) \cite{Susskind_1995,Hooft_2001} in a framework termed \textit{Fractional Holographic Dark Energy} (FHDE) (see \cite{Trivedi_2024, Bidlan_2025} for a detailed review). The initial research on the Holographic Dark Energy (HDE) framework originated from a quantum field theory (QFT) approach that links a short-distance cutoff to a long-distance cutoff due to the constraints imposed by black hole formation, specifically Cohen's inequality \cite{Cohen_1999}. Specifically, if $\rho$ represents the quantum zero-point energy density from a short-distance cutoff, then the total energy within a region of size $L$ should not exceed the mass of a black hole of the same size, leading to the inequality $L^{3}\rho\leq LM^{2}_{pl}$. Therefore, the maximum allowable values for the infrared (IR) cutoff $L_{\text{IR}}$ satisfy this inequality. In conventional HDE, the expression of density is given as \cite{Wang_2017}:
\begin{equation}
    \rho=3c^{2}L^{-2}_{\text{IR}}
\end{equation}
where $c$ is an arbitrary positive parameter, and we set $M_{pl}=1$ (Planck's mass). This framework has found extensive application in cosmology, particularly in explaining the late-time expansion of the Universe \cite{Li_2013, Wang_2017, Trivedi_2024, Bidlan_2025, Trivedi:2025sbe, Nojiri_2017, Nojiri_2006, Trivedirip_2024, TrivediScherrer_2024}. Moreover, numerous other studies have extended the HDE framework based on modified Bekenstein-Hawking entropy and have also highlighted its own challenges in \cite{Tavayef_2018, Oliveros_2022, Saridakis_2020, Zhang_2009, Colgain:2021beg}. Within the framework of FHDE \cite{Trivedi_2024}, the extension comes in due to modification of the underlying mathematical structure, i.e., replacing conventional calculus tools, namely, derivatives by fractional integro-differential operators (Riesz derivation \cite{Jalalzadeh_2021}) in obtaining a quantum fractionally extended form of the Bekenstein-Hawking horizon entropy in \cite{Jalalzadeh_2021}, written as:
\begin{equation}\label{fractional entropy}
    S_{\text{horizon}}\propto A^{\frac{\alpha+2}{2\alpha}}
\end{equation}
Here $\alpha$ is termed as Lévy's index or simply a ``fractional parameter", and it is allowed to take values within the range of $1<\alpha\leq2$. In our recent studies, 
we examined FHDE within the Hubble cutoff, $ L=H^{-1}$, and alleviated a few issues faced in conventional HDE \cite{Wang_2017}. Namely, the issue of ``constancy", i.e., no dynamic response of dark energy in the Hubble cutoff, was addressed by including a fractionally extended HDE framework, i.e., FHDE. Additionally, we provided a comprehensive study, corroborated by plots of relevant cosmological parameters in ref.  \cite{Trivedi_2024}. Moreover, we further extended the FHDE perspective to reconstruct it through various bosonic/string-inspired and one-gauge-boson-condensate dynamical effective field configurations, as discussed in ref. \cite{Bidlan_2025}. Looking forward within the context of this paper we are interested in examining whether future rip scenarios such as the big rip, little rip, or pseudo rip can emerge naturally within the FHDE framework. Let us first highlight the expression of FHDE density obtained after substituting the quantum fractionally modified horizon entropy in Cohen's inequality \cite{Cohen_1999}, which can be written as:
\begin{equation}\label{FHDE density}
    \rho_{\text{de}}=3c^{2}L^{\frac{2-3\alpha}{\alpha}}
\end{equation}
Quite evidently, the conventional HDE scenario is obtained as the fractional features start to diminish in the limit $\alpha\rightarrow2$. Although our primary interest is in extracting novel insights from late-time cosmology when fractional features dominate, i.e., when $\alpha=1.1$ or anywhere within the range $1<\alpha\leq2$. Moreover, in recent literature, the authors in ref. \cite{TrivediScherrer_2024, Brevik:2024ozg} presented intriguing results in investigating future rips and other kinds of singularities, such as the \textit{Big Freeze}, within the conventional, Barrow, and Tsallis HDE frameworks through an ansatz-based approach which involved employing the ansatz for $H(t)$ that gives rise to little and pseudo rip singularities. To clarify, within our work, we will not address the big freeze singularity, as it presents its own challenges and requires a separate study, which we defer to future work. Furthermore, within the scope of this work, we will extend the work in ref. \cite{Trivedirip_2024} and \cite{TrivediScherrer_2024} on rip scenarios through the use of FHDE, placing special emphasis on the non-local features induced by FC and investigating how they contribute to the occurrence or avoidance of these late-time events within the FHDE framework. 

\bigskip

In particular, a substantial body of literature has emerged, focusing on the exploration of various types of singularities in classical and quantum cosmology that may arise in the future evolution of the Universe \cite{Albarran_2016,Borislavov_Vasilev_2021,Bouhmadi-Lopez:2017ckh,D_browski_2006,Albarran:2018mpg,Albarran:2015tga}. The detection of late-time acceleration has significantly propelled such investigations intensely \cite{Trivedirip_2024, Trivedi_2024sym, Pati_2022}. A particularly intriguing class of such far-future events is rip scenarios, where the Universe proceeds toward progressive disintegration in various capacities. A few interesting questions within the framework 
of FHDE are to then ask: ``\textit{Which rip scenarios could possibly occur within the FHDE model?}" and ``\textit{Could the preference of non-local features avoid any of the rip scenarios?}". By ``non-local features," we mean a possible consequence of the memory effect in the cosmological evolution of dark energy during late times. Within the FHDE framework with a Lévy index $\alpha$, the memory kernel of a fractional derivative encodes how strongly past cosmic epochs influence the current behaviour of dark energy \cite{Landim_2021,Rasouli_2024,Micolta-Riascos:2023mqo}. This is particularly relevant for rip singularities, as the non-local terms can either enhance or suppress singularity formation depending on how memory accumulates during the approach to finite-time futures. A recent work in ref. \cite{TrivediScherrer_2024} investigates the occurrence of all the rip singularities in the context of non-fractional HDE models. In particular, the authors employ the GO cutoff \cite{Granda_2008}, and find that the overarching possibility is that of the big and pseudo rip, while the little rip is likely to occur for very special class of IR cutoffs \cite{TrivediScherrer_2024}. 
We will investigate the above-mentioned questions within the FHDE framework in the next sections. For the sake of completeness and the reader, we highlight the key distinguishing features of several rip scenarios as follows:

\begin{itemize}
    \item \textbf{Big Rip (Type I Singularity): }A late-time cosmic event where $t\rightarrow t_{\text{finite}}$, both the effective energy density and pressure of the Universe diverges, i.e., $\rho_{\text{eff}}\rightarrow\infty$ and $p_{\text{eff}}\rightarrow-\infty$, and the Hubble parameter also diverges, $H(t)\rightarrow\infty$ \cite{Caldwell_2003}. The expression for the Hubble parameter leading to the big rip singularity can be expressed as:
    \begin{equation}\label{big rip ansatz}
        H(t)\approx \frac{H_{0}}{(t_{\text{rip}}-t)^{m}}
    \end{equation}
    Here $m$ is a positive arbitrary constant, and we will take $m=1$ throughout this paper. In a late-time big rip singularity, the dark energy EoS parameter $w$ remains phantom, satisfying $w<-1$, with a constant value or approaching a limit strictly below $-1$ as the singularity is reached in finite time.

    \item \textbf{Little Rip: }In this late-time scenario, the effective energy density, pressure, and the Hubble parameter asymptotically tend to infinite values as $t\rightarrow\infty$. Therefore, in sharp contrast to the Big Rip singularity, the Little Rip singularity is not a finite-time singularity \cite{Frampton_2011}. Here, the expression for the Hubble parameter can be expressed as:
    \begin{equation}\label{little rip ansatz}
        H(t)\approx H_{0}\exp{(\lambda t)}
    \end{equation}
    In a little rip, the dark energy EoS parameter remains phantom throughout late times, with the energy density and Hubble parameter diverging, unlike the finite time big rip.

    \item \textbf{Pseudo Rip: }Here, the Hubble parameter increases monotonically with $t\rightarrow\infty$, but it is bounded from above by the value $ H_{\infty}\equiv H_{t=\infty}$ so that $H(t)\rightarrow H_{\infty}$ as $t\rightarrow\infty$. All the bound structures are only partially disrupted before the Universe asymptotes to de Sitter \cite{Frampton_2012}. For this scenario, the Hubble parameter can be written as:
    \begin{equation}\label{pseudo rip ansatz}
        H(t)\approx H_{0}-H_{1}\exp{(-\lambda t)}
    \end{equation}
    In a pseudo rip, the dark energy EoS parameter stays phantom, $w_{\text{de}}(t)<-1$, for the whole cosmic history but approaches $w_{\text{de}}(t)\rightarrow-1$ from below at late times, $w_{\text{de}}(t)\rightarrow-1^{-}$, so that the expansion asymptotically tends to de Sitter rather than diverging like a little rip. By definition, pseudo rips are ``intermediate" between a constant cosmological constant ($w_{\text{de}}=-1$) and a little rip ($w_{\text{de}}<-1$ driving $\rho_{\text{de}}\rightarrow+\infty$ as $t\rightarrow\infty$) \cite{Frampton_2012}.
\end{itemize}

 We begin our investigation with a review of the cosmology 
 describing an interacting FHDE framework in Section \ref{Section 2}. In Section \ref{Section 3}, we investigate whether all the rip scenarios, i.e., big, little, and pseudo rip, can occur naturally with the GO cutoff within the interacting FHDE setting. In Section \ref{Section 4}, our investigation shifts to an ansatz-based approach, where we employ some well-explored ansatz in literature for the Hubble parameter, $H(t)$, and examine how it defines the nature of key cosmological parameters such as the Equation of State (EoS) and the Squared Sound Speed (SSS) parameter for little and pseudo-rip within the Hubble cutoff. And finally, we conclude our work in Section \ref{Section 5} and outline possible future research directions in the FHDE framework.

\section{Interacting Fractional HDE Framework}\label{Section 2}

In this section, we will establish the important expressions of the key parameters that will play a crucial role in our investigation of future rip scenarios within FHDE. We will be working within the interacting sector as mentioned earlier. One can then begin by establishing the Friedmann equation for the interacting dark sector in a flat ($k=0$) FRW Universe. By using the definitions: $\Omega_{\text{de}}=\rho_{\text{de}}/3H^{2}$ and $\Omega_{\text{dm}}=\rho_{\text{dm}}/3H^{2}$, one can write
\begin{equation}\label{friedmann}
    \Omega_{\text{de}}+\Omega_{\text{dm}}=1
\end{equation}
The subscripts ``de" and ``dm" stand for dark energy and dark matter, respectively. The expression for the dark energy density parameter $\Omega_{\text{de}}$ can be written as:
\begin{equation}
\Omega_{\text{de}}=\frac{\rho_{\text{de}}}{3H^{2}}=c^{2}\frac{L^{\frac{2-3\alpha}{\alpha}}}{H^{2}}
\end{equation}
One can then always find the expression for the dark matter density $\Omega_{\text{dm}}$ using the Friedmann equation established in Eq. (\ref{friedmann}). Moving ahead, we require the continuity equations that encapsulate interactions between dark energy and dark matter through means of a term $Q$ that sits on the RHS of the continuity equation. The primary reason for incorporating interaction between dark energy and dark matter is to understand how the coupling between dark energy and dark matter can amplify or improve the occurrence of the rip scenario that we are interested in\footnote{For an impatient reader, we highlight that the $Q$ term plays a key role in producing a big rip within the Granda-Oliveros cutoff.}. The continuity equations are written as
\begin{equation}\label{continuityeqn1}
\dot{\rho}_{\text{de}}+3H\rho_{\text{de}}(1+w_{\text{de}})=-Q
\end{equation}
\begin{equation}\label{continuityeqn2}
    \dot{\rho}_{\text{dm}}+3H\rho_{\text{dm}}(1+w_{\text{dm}})=+Q
\end{equation}
Here, we consider two cases for the $Q$ terms. We consider the following expressions for the linear and non-linear interactions, namely:
\begin{itemize}
    \item Linear interaction term can be written as:
    \begin{equation}\label{linear Q}
        Q=9H^{3}\beta(\Omega_{\text{de}}+\Omega_{\text{dm}})
    \end{equation}

    \item Non-linear interaction term can be written as:
    \begin{equation}\label{nonlinear Q}
        Q=3H\beta\left(\frac{\Omega_{\text{de}}}{1-\Omega_{\text{de}}}\right)
    \end{equation}
\end{itemize}
Here $\beta$ represents the strength of the interaction between dark energy and dark matter densities. It is commonly termed as ``coupling constant". It is normally constrained to be \textit{very} small, i.e., $\beta\approx0$, but we will highlight how a slight change in its value can influence the occurrence of a few rip scenarios in Section \ref{Section 3} and \ref{Section 4}. Furthermore, now we are ready to set up the FHDE scenario by writing the expression for the time derivative of its dark energy density (see Eq. (\ref{FHDE density})) as:
\begin{equation}
    \dot{\rho}_{\text{de}}=\rho_{\text{de}}\left(\frac{2-3\alpha}{\alpha}\right)\frac{\dot{L}}{L}
\end{equation}
Putting this into the continuity Eq. (\ref{continuityeqn1}), we obtain the expression for the EoS parameter $w_{\text{de}}(t)$ in terms of a general IR cutoff $L$ as follows:
\begin{equation}\label{general EoS}
    w_{\text{de}}(t)=-1-\frac{1}{3H\Omega_{\text{de}}}\left[\frac{Q}{3H^{2}}+\Omega_{\text{de}}\left(\frac{2-3\alpha}{\alpha}\right)\frac{\dot{L}}{L}\right]
\end{equation}
In addition, we also write down the expression for the squared sound speed parameter $v^{2}_{s}(t)$ as:
\begin{equation}\label{general vs}
    v^{2}_{s}(t)=w_{de}+\dot{w}_{\text{de}}\left(\frac{\alpha}{2-3\alpha}\right)\frac{L}{\dot{L}}
\end{equation}
In the next section, we will investigate all the rip scenarios, which are famously a consequence of the phantom energy that dominates over all other forms of matter during the late-time expansion of the Universe. In particular, the subsequent sections will investigate these rip scenarios, corroborated with numerical plots of the parameters defined in Eqs. (\ref{general EoS}) and (\ref{general vs}) for Granda-Oliveros and Hubble cutoff. 

\section{Rip Scenarios in Granda-Oliveros Cutoff}\label{Section 3}

We write down the Friedmann equation during the dark energy domination, i.e., when the Universe transitions into the late-time accelerated expansion, as:
\begin{equation}\label{Friedmann}
    H^{2}=\frac{\rho_{\text{de}}}{3}= c^{2}L^{\frac{2-3\alpha}{\alpha}}
\end{equation}
In this section, the choice of $L$ is made such that all the rip scenarios can be investigated. This choice happens to be the Granda-Oliveros cutoff because it provides us with the essential quantities, such as the time derivative of $H(t)$, which allows for a more thorough investigation. This choice has been well-explored within the literature due to its applicability in various late-time cosmological scenarios \cite{Trivedirip_2024, TrivediScherrer_2024, Granda_2008}. Note that, in general, HDE models cannot be treated with any of the parametrisations discussed in ref. \cite{Frampton_2011, Frampton_2012}, because the standard Friedmann equation does not apply to them. This will lead to a different parametrisation that can be examined in the context of the little rip and big rip models. Consider the choice for the IR cut-off scale $L$. An early suggestion was to consider a cutoff scale given by $L\rightarrow H^{-1}$. This choice aimed to alleviate the fine-tuning problem by introducing a natural length scale associated with the inverse of the Hubble parameter $H$. However, it was found that this particular scale resulted in an equation of state approaching zero, failing to contribute significantly to the current accelerated expansion of the universe. An alternative approach involved utilising the particle horizon as the length scale. This alternative resulted in an equation of state parameter higher than $-1/3$. However, despite this modification, the challenge of explaining the present acceleration remained unresolved. Another option considered the future event horizon as the length scale. Although the desired acceleration regime can be achieved in this case, this approach raises problems with causality, posing a significant obstacle to its viability. To circumvent these difficulties, the Granda-Oliveros cutoff was proposed in Ref. \cite{Granda_2008}. This cutoff is defined in the following manner:
\begin{equation}\label{G-O}
    L=(\gamma H^{2}+\delta\dot{H})^{-\frac{1}{2}}
\end{equation}
The corresponding FHDE density expression becomes:
\begin{equation}
    \rho_{\text{GO}}=3c^{2}(\gamma H^{2}+\delta\dot{H})^{\frac{3\alpha-2}{2\alpha}}
\end{equation}
In addition, we also require the expression of $\Omega_{\text{GO}}=\rho_{\text{GO}}/3H^{2}$, which can be written as:
\begin{equation}
    \Omega_{\text{GO}}=c^{2}\frac{(\gamma H^{2}+\delta\dot{H})^{\frac{3\alpha-2}{2\alpha}}}{H^{2}}
\end{equation}
Plugging the cutoff (\ref{G-O}) into the Friedmann equation (\ref{Friedmann}) allows us to write:
\begin{equation}\label{Hdot}
    \dot{H}=\frac{1}{\delta}\left[\left(\frac{H}{c}\right)^{n}-\gamma H^{2}\right],\quad n=\frac{4\alpha}{3\alpha-2}
\end{equation}
so,
\begin{equation}\label{Hint}
    \int_{H_{i}}^{H_{f}}\frac{\delta}{\left(H/c\right)^{n}-\gamma H^{2}}dH=\int_{t_{i}}^{t_{f}}dt
\end{equation}
Here $c$, $\gamma$, and $\delta$ are all positive constants. The time derivative of the Hubble parameter described in Eq. (\ref{Hdot}) gives two solutions: positive ($\dot{H}>0$) and negative ($\dot{H}<0$). Within our investigation of a rip scenario, one can safely reject the negative solution, as it does not yield a rip scenario for $\dot{H}<0$. Therefore, we focus on the positive solution, i.e., $\dot{H}>0$, which holds insights into different kinds of interesting rip scenarios. Let us highlight that the big rip again occurs if the solution $H(t)$ diverges at the finite cosmic time $t_{f}$, while a pseudo rip occurs if $H(t)$ monotonically increases but tends to a finite constant $H_{\infty}$ as $t\rightarrow\infty$. In the asymptotic limit of Eq. (\ref{Hdot}), the sign and power of the dominant term control whether $H(t)$ blows up, saturates, or decays. Let us examine Eq. (\ref{Hint}) for the possibility of the big rip ($H\rightarrow\infty$ at finite $t_{f}$). The integral will converge whenever $n\geq2$, indicating a \textit{big rip} singularity at a finite time. For $n<2$, $H(t)$ goes to a
finite value as $t\rightarrow\infty$, corresponding to a \textit{pseudo-rip}. The condition $n\geq2$ translates to the occurrence of the big rip condition in terms of the Lévy's index as $\alpha\leq2$, whereas the pseudo rip condition for $\alpha$ becomes $\alpha>2$. It is worth noting that $\alpha$ is allowed to take values in the range $1<\alpha\leq2$, indicating that pseudo rip does not stand well against our fractional HDE approach with a Granda-Oliveros cutoff (\ref{G-O}), whereas for the case of big rip, we conclude that it is a possible late time event within the established cosmology. In particular, one can then say that within the fractional construction, i.e., FHDE \cite{Trivedi_2024, Bidlan_2025}, the fractional values of $\alpha$ for the occurrence of big rip singularity highlight the existence of non-local features (or memory effect) playing their role during late-time accelerated expansion.

\bigskip

So far, we have examined the FHDE model, which results in the occurrence and avoidance of big and pseudo rips, respectively, in the FHDE framework. We now demonstrate that FHDE models with Granda-Oliveros cutoff fail to produce a little rip singularity except for very special cases of the cutoff. In fact, a similar analysis was presented in Ref. \cite{TrivediScherrer_2024} for conventional HDE, where the authors set the expression for dark energy density as a free function: $\rho_{\Lambda}=3c^{2}[f(L)]^{-2\mathcal{A}}$. Here $\mathcal{A}=(3\alpha-2)/2\alpha$. Now, plugging this expression of energy density into the Friedmann equation (\ref{Friedmann}), we get: $\left(\frac{H}{c}\right)^{-\frac{1}{\mathcal{A}}}=f(L)$. Within this section, we have set $L$ to describe the Granda-Oliveros cutoff (\ref{G-O}). This choice leads to the following integral relation between $H$ and $t$:
\begin{equation}
    \int_{H_{i}}^{H_{f}}\frac{\delta}{\left\{f^{-1}\left[\left(\frac{H}{c}\right)^{-\frac{1}{\mathcal{A}}}\right]\right\}^{-2}-\gamma H^{2}}dH=\int_{t_{i}}^{t_{f}}dt
\end{equation}
The one possibility of a little rip corresponds to the case where the quantity $\left\{f^{-1}\left[\left(\frac{H}{c}\right)^{-\frac{1}{\mathcal{A}}}\right]\right\}^{-2}\sim\gamma H^{2}+g(H)$ as $H(t\rightarrow\infty)\rightarrow\infty$, where $\int dH/g(H)$ diverges. Notably, the form of $L\sim(\gamma H^{2}+g(H))^{-\frac{1}{2}}$ belongs to a larger category of cutoffs, namely, the Nojiri-Odintsov (NO) cutoff \cite{Nojiri_2017, Nojiri_2006}. However, as it is evident, such behaviour is rather contrived and not well-established in the literature of IR cutoffs for the HDE framework. Therefore, we conclude that the little rip is not a possible future evolution except for a very special class of cutoffs within the FHDE framework, as shown above. Therefore, one struggles to motivate a particular form for the cutoff $L$ even if one wants to allow for the generality of the cutoff (see \cite{TrivediScherrer_2024}).

\bigskip

\begin{figure*}
    \centering
    \begin{subfigure}[b]{0.45\linewidth}
        \centering
        \includegraphics[width=\linewidth]{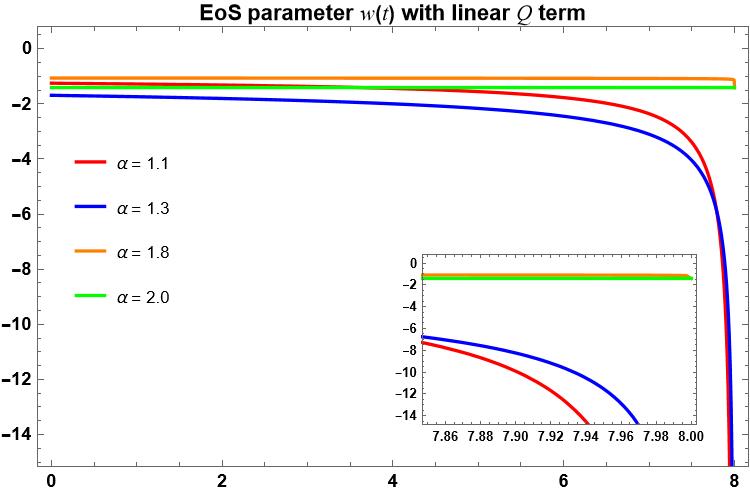}
        \caption{Plot for EoS parameter $w_{\text{GO}}(t)$ against cosmic time $t$ for various values of Lévy's index $\alpha$ within an ansatz established in Eq. (\ref{big rip ansatz}) and linear Q term for big rip.}
        \label{Figure 1a}
    \end{subfigure}
    \hfill
    \begin{subfigure}[b]{0.45\linewidth}
        \centering
        \includegraphics[width=\linewidth]{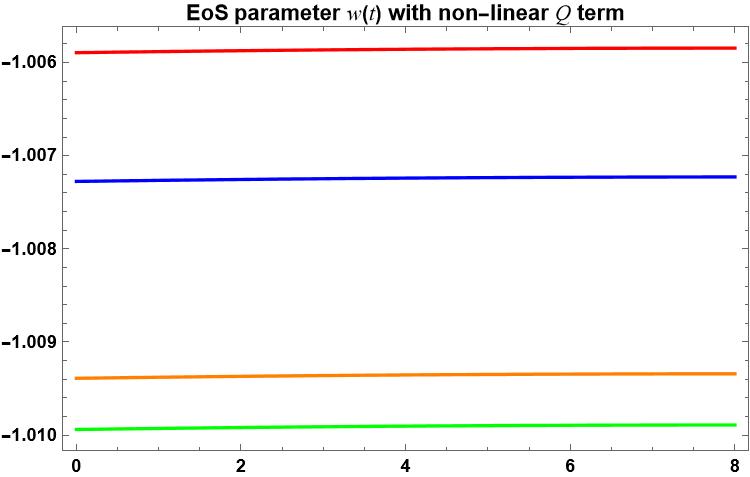}
        \caption{Plot for EoS parameter $w_{\text{GO}}(t)$ against cosmic time $t$ for various values of Lévy's index $\alpha$ within an ansatz established in Eq. (\ref{big rip ansatz}) and non-linear $Q$ term for big rip.}
        \label{Figure 1b}
    \end{subfigure}
    \caption{Plots for EoS parameter $w_{\text{GO}}(t)$ for big rip within Granda-Oliveros cutoff.}
    \label{Figure 1}
\end{figure*}

\begin{figure*}
    \centering
    \begin{subfigure}[b]{0.45\linewidth}
        \centering
        \includegraphics[width=\linewidth]{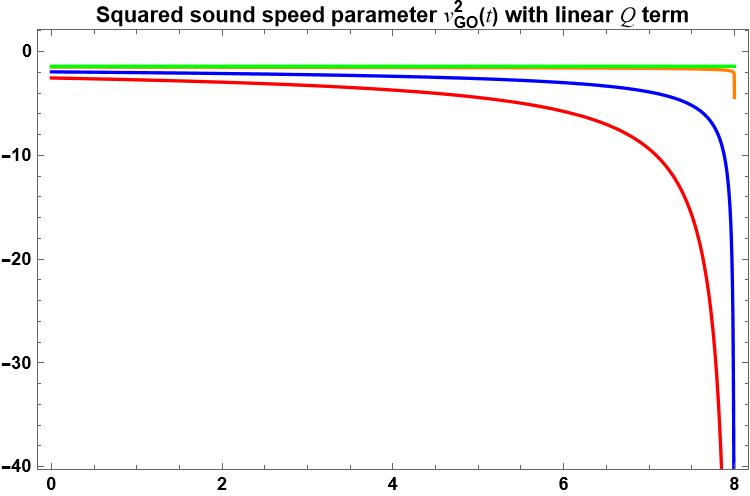}
        \caption{Plot for squared sound speed parameter $v^{2}_{\text{GO}}(t)$ against cosmic time $t$ for various values of Lévy's index $\alpha$ within an ansatz established in Eq. (\ref{big rip ansatz}) and linear Q term for big rip.}
        \label{Figure 2a}
    \end{subfigure}
    \hfill
    \begin{subfigure}[b]{0.45\linewidth}
        \centering
        \includegraphics[width=\linewidth]{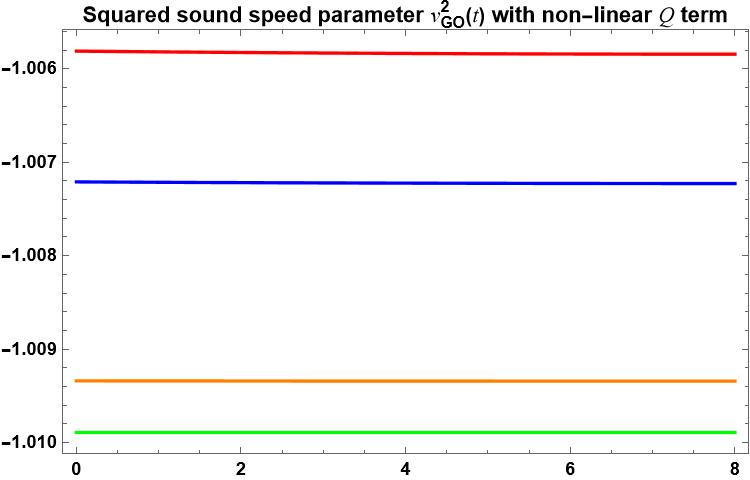}
        \caption{Plot for squared sound speed parameter $v^{2}_{\text{GO}}(t)$ for various values of Lévy's index $\alpha$ within an ansatz established in Eq. (\ref{big rip ansatz}) and non-linear $Q$ term for big rip.}
        \label{Figure 2b}
    \end{subfigure}
    \caption{Plots for squared sound speed parameter $v^{2}_{\text{GO}}(t)$ for big rip within Granda-Oliveros cutoff.}
    \label{Figure 2}
\end{figure*}

Moreover, within the FHDE density for the Granda-Oliveros cutoff, one can estimate the EoS parameter as a function of time $t$ by utilising the interacting continuity equation (\ref{continuityeqn1}) with the following expression for $\dot{\rho}_{\text{de}}$ as:
\begin{equation}\label{rhodot-GO}
    \dot{\rho}_{\text{GO}}=\mathcal{A}\rho_{\text{GO}}\left(\frac{2\gamma H\dot{H}+\delta\Ddot{H}}{\gamma H^{2}+\delta\dot{H}}\right)
\end{equation}
Here $\mathcal{A}=(3\alpha-2)/2\alpha$. We put this expression of $\dot{\rho}_{\text{de}}$ (\ref{rhodot-GO}) in Eq. (\ref{continuityeqn1}) and get:
\begin{equation}
\begin{split}
    w_{\text{GO}}(t)=-1-\frac{1}{3H}\Bigg[\mathcal{A}\hspace{0.2mm}\left(\frac{2H\dot{H}\gamma+\delta\Ddot{H}}{\gamma H^{2}+\delta\dot{H}}\right)+\frac{Q}{3H^{2}\Omega_{\text{GO}}}\Bigg]
\end{split}
\end{equation}
 For big rip, the Hubble parameter must obey such behaviour as expressed in Eq. (\ref{big rip ansatz}), i.e., $H(t)\approx H_{0}(t_{\text{rip}}-t)^{-m}$. Here, $m>0$ represents a positive arbitrary constant, and $t_{\text{rip}}$ represents the comoving rip singularity time. In the limiting case $t\rightarrow t_{\text{rip}}$, the Hubble parameter starts to diverge for any positive value of $m$. In addition, one can then also derive the expression for the squared sound speed parameter within the Granda-Oliveros cutoff, $v^{2}_{\text{GO}}$, written as
\begin{equation}
    v^{2}_{\text{GO}}=w_{\text{GO}}+\frac{\dot{w}_{\text{GO}}}{\mathcal{A}}\left(\frac{\gamma H^{2}+\delta\dot{H}}{2\gamma H\dot{H}+\delta\Ddot{H}}\right)
\end{equation}
Now that we have the analytic expressions for the EoS and squared sound speed parameters within the Granda-Oliveros cutoff, we can produce some interesting plots depicting their late-time evolution. In particular, Figures (\ref{Figure 1}) and (\ref{Figure 2}) showcase the cosmological evolution of EoS and squared sound speed parameter for linear and non-linear interaction term $Q$. While plotting for these parameters, we set $\beta=0.01$ (weakly interacting)\footnote{While for $\beta=0$, i.e., non-interacting scenario, no big rip occurs. On the other hand, as coupling strength grows, i.e., $\beta\rightarrow1$, the big rip becomes more pronounced. Nevertheless, we consider $\beta=0.01$ within our considerations, as this yields a plausible result.}, $c=0.495$\footnote{See \cite{Li_2013} for Planck constraints on HDE parameters. In particular, we set $c=0.495$ obtained from Planck+WP+BAO+HST+lensing \cite{Li_2013}.}, $H_{0}=67.6$\footnote{See \cite{guo2024newestmeasurementshubbleconstant} for recent constraints on Hubble parameter.}, $\gamma=0.9$ and $\delta=0.5$ \cite{10.1093/mnras/stae2257} for various values of $\alpha$ within the range $1<\alpha\leq2$. Let us now elaborate on these plots as follows:
\begin{itemize}
    \item Let us first consider the linearly interacting scenarios as shown in Figure (\ref{Figure 1a}) and (\ref{Figure 2a}):
    \begin{itemize}
        \item In particular, for fractional values of $\alpha$, such as $\alpha=1.1$ and $1.3$, the EoS parameter tends to negative infinity crossing the phantom divide ($w<-1$) in the late-time limit. This is a well-anticipated result because one expects the EoS parameter to become infinitely negative as the Universe asymptotically approaches a big rip singularity.

        \item While for higher values of $\alpha$, i.e., $1.8$ and $2.0$, the EoS parameter hovers around $w\sim-1$, indicating the absence or weakening of big rip-like singularity as fractional features become scarce in the limit $\alpha\rightarrow2$.

        \item On the other hand, the squared sound speed (SSS) parameter remains negative for all values of $\alpha$. It becomes more negative as the fractional feature begins to dominate $\alpha\rightarrow1.1$, indicating the classical instability during the big rip singularity, which is again a well-anticipated behaviour due to the excess of phantom energy.
    \end{itemize}

\item Let us now consider the non-linearly interacting scenarios as shown in Figure (\ref{Figure 1b}) and (\ref{Figure 2b}):

\begin{itemize}
    \item The EoS parameter and the SSS parameter for all values of $\alpha$ hover near $-1$ for both the parameters, which corresponds to a nearly de Sitter or very mild phantom evolution, not a genuine big-rip type finite-time divergence.

    \item With a non-linear $Q$ term, the cosmological evolution of EoS and SSS parameter is entirely opposite to that of the evolution described by the same parameters with a linear $Q$ term in the late-time limit.

    \item Therefore, the big rip is only observed when the $Q$ term is linear in nature, with $\alpha$ taking values near $\alpha\rightarrow1.1$.
\end{itemize}
\end{itemize}
In essence, for the linear interaction case, fractional values $\alpha\rightarrow1.1$ drive the EoS parameter $w_{\text{HH}}(t)$ to large negative values with $v_{\text{HH}}^{2}<0$, leading to a classically unstable big-rip–type evolution. For larger $\alpha$ and for all non-linear interaction cases, both $w_{\text{HH}}(t)$ and $v_{\text{HH}}^{2}<0$ remain close to $-1$ without finite-time divergences, corresponding instead to a mild, de Sitter phantom behaviour.


\section{Little and Pseudo Rip in Hubble Cutoff}\label{Section 4}

Before we proceed, we would like to highlight the reason for our investigation into the existence of late-time cosmological singularities within the Hubble cutoff using an ansatz-based approach in the FHDE framework. In recent literature \cite{Trivedirip_2024} and \cite{TrivediScherrer_2024}, physicists have developed a methodology for investigating the possibility of finding a small rip and pseudo-rip within conventional HDE, as well as other extensions, such as Barrow and Tsallis HDE. Within the boundaries of our work, earlier in \cite{Trivedi_2024}, we showed that a dynamic nature of dark energy can be predicted within the Hubble cutoff, i.e., $L=H^{-1}$, which is \textit{not} the case in the conventional HDE scenario. Therefore, extending with this as our primary motivation, we intend to import the well-explored \textit{ansatz} for the Hubble parameter $H(t)$ in the literature for little rip and pseudo rip to explore some key parameters that will help us understand the very nature of little and pseudo rip within the Hubble cutoff.

\bigskip

\begin{figure*}
    \centering
    \begin{subfigure}[b]{0.45\linewidth}
        \centering
        \includegraphics[width=\linewidth]{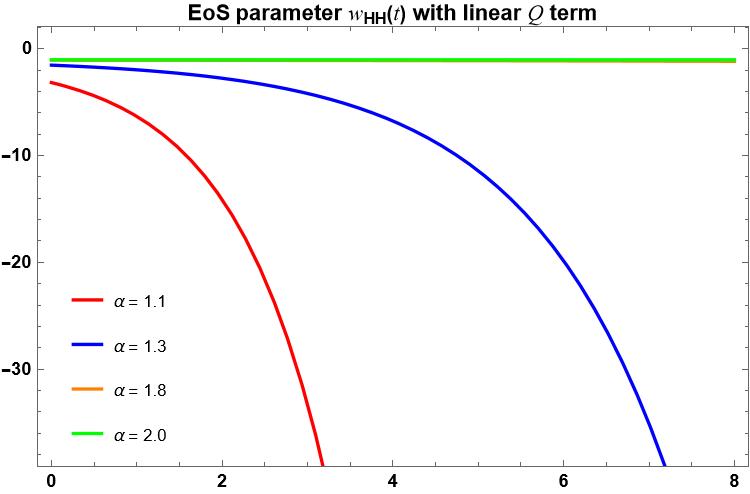}
        \caption{Plot for EoS parameter $w_{\text{HH}}(t)$ against cosmic time $t$ for various values of Lévy's index $\alpha$ within an ansatz established in Eq. (\ref{little rip ansatz}) and $Q$ term in Eq. (\ref{linear Q}) for little rip.}
        \label{Figure 3a}
    \end{subfigure}
    \hfill
    \begin{subfigure}[b]{0.45\linewidth}
        \centering
        \includegraphics[width=\linewidth]{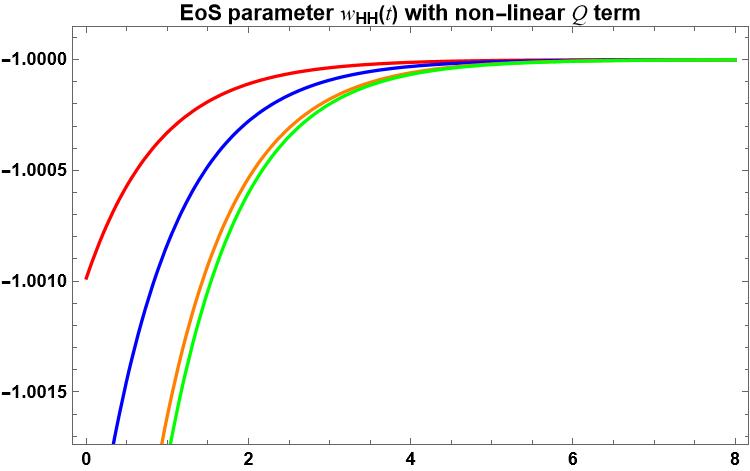}
        \caption{Plot for EoS parameter $w_{\text{HH}}(t)$ against cosmic time $t$ for various values of Lévy's index $\alpha$ within an ansatz established in Eq. (\ref{little rip ansatz}) and $Q$ term in Eq. (\ref{nonlinear Q}) for little rip.}
        \label{Figure 3b}
    \end{subfigure}
    \caption{Plots for EoS parameter $w_{\text{HH}}(t)$ for little rip within Hubble cutoff.}
    \label{Figure 3}
\end{figure*}

\begin{figure*}
    \centering
    \begin{subfigure}[b]{0.45\linewidth}
        \centering
    \includegraphics[width=1\linewidth]{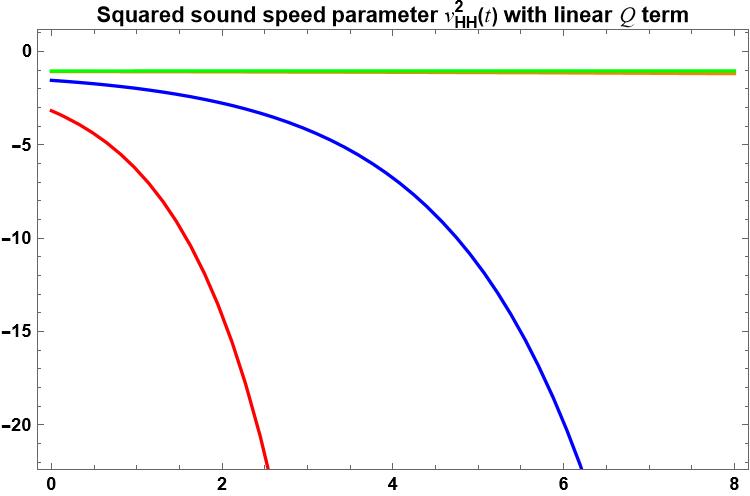}
    \caption{Plot for squared sound speed parameter $v^{2}_{\text{HH}}$ against cosmic time $t$ for various values of Lévy's index $\alpha$ within an ansatz established in Eq. (\ref{little rip ansatz}) and $Q$ term in Eq. (\ref{linear Q}) for little rip.}
    \label{Figure 4a}
    \end{subfigure}
    \hfill
    \begin{subfigure}[b]{0.45\linewidth}
        \centering
    \includegraphics[width=1\linewidth]{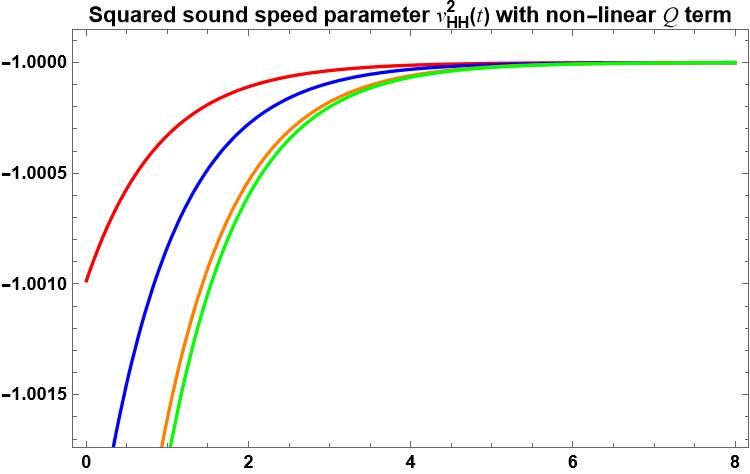}
    \caption{Plot for squared sound speed parameter $v^{2}_{\text{HH}}$ against cosmic time $t$ for various values of Lévy's index $\alpha$ within an ansatz established in Eq. (\ref{little rip ansatz}) and $Q$ term in Eq. (\ref{nonlinear Q}) for a little rip.}
    \label{Figure 4b}
    \end{subfigure}
    \caption{Plots for squared sound speed parameter $v^{2}_{\text{HH}}$ for little rip within Hubble cutoff.}
    \label{Figure 4}
\end{figure*}

The expression for the time derivative of the FHDE density within the Hubble cutoff can be written as:
\begin{equation}
    \dot{\rho}_{\text{HH}}=2\mathcal{\mathcal{A}}\frac{\rho_{\text{HH}}\dot{H}}{H}
\end{equation}
Here, $\mathcal{A}=(3\alpha-2)/2\alpha$. We first examine the FHDE framework for the little rip ansatz established in Eq. (\ref{little rip ansatz}) with a general form of $Q$ term. The following expression of the EoS parameter $w_{\text{HH}}(t)$ is defined within the Hubble cutoff $L=H^{-1}$:
\begin{equation}\label{EoSHH}
\begin{split}
    w_{\text{HH}}(t)=-1-\frac{1}{3H\Omega_{\text{de}}}\Bigg[\frac{Q}{3H^{2}}+2\mathcal{A}\hspace{1mm}\frac{\Omega_{\text{de}}\dot{H}}{H}\Bigg]
\end{split}
\end{equation}
In addition, using Eq. (\ref{general vs}) with Eq. (\ref{EoSHH}), we get:
\begin{equation}\label{vsHH}
    v^{2}_{\text{HH}}(t)=w_{\text{HH}}+\frac{\dot{w}_{\text{HH}}}{2\mathcal{A}}\frac{H}{\dot{H}}
\end{equation}
The above written expression represents the SSS parameter within the Hubble cutoff. Moreover, using these analytic expressions for the EoS parameter (\ref{EoSHH}) and the SSS parameter (\ref{vsHH}), we present some interesting plots depicting their late-time evolution for the little rip ansatz (\ref{little rip ansatz}). We plot them for both cases of $Q$, i.e., linear and non-linear interactions within the dark sector in Figures (\ref{Figure 3}) and (\ref{Figure 4}) and summarise our findings as follows:

\begin{itemize}

\item Let us first elaborate on the linearly interacting scenario (\ref{linear Q}) in Figures (\ref{Figure 3a}) and (\ref{Figure 4a}):

\begin{itemize}

    \item In Figure (\ref{Figure 3a}), we observe that the EoS parameter within the Hubble cutoff, $w_{\text{HH}}(t)$ (\ref{EoSHH}), for a linear $Q$ term tends to diverge towards negative infinity for fractional values of $\alpha$, such as $1.1$ and $1.3$ in late-time limit. This behaviour is quite analogous to that of the big rip in Figure (\ref{Figure 1a}).

    \item On the other hand, for higher values of $\alpha\rightarrow2$, the EoS parameter remains constant even during late-time expansion, indicating no little rip singularity-like behaviour. 

    \item The corresponding SSS parameter within the Hubble cutoff, $v^{2}_{\text{HH}}$, in Figure (\ref{Figure 2a}), remains negative throughout the cosmic history and the late time expansion, indicating classical instability for the ansatz (\ref{little rip ansatz}) in FHDE.

    \item  Therefore, the EoS and SSS parameters diverge to large negative values at a finite time, which corresponds to a big‑rip–type finite‑time singularity rather than a genuine little rip.
    
\end{itemize}

\item Let us now elaborate on the non-linearly interacting scenario (\ref{nonlinear Q}) in Figures (\ref{Figure 3b}) and (\ref{Figure 4b}):

\begin{itemize}
    \item In Figure (\ref{Figure 3b}), we observe that the EoS parameter within the Hubble cutoff $w_{\text{HH}}(t)$ (\ref{EoSHH}) for a linear $Q$ term tends to asymptote towards $-1$ from below for all fractional values of $\alpha$, which is a well suited behavior for a little rip during late-time.

    \item Notably, for higher values of $\alpha\rightarrow2$, the EoS parameter asymptotes towards $-1$, but the evolution is less dynamic, i.e., slower, compared to that of evolution governed by $\alpha$ near $\alpha\rightarrow1.1$ region.

    \item The corresponding SSS parameter within the Hubble cutoff, $v^{2}_{\text{HH}}$, remains negative through the cosmic history and late time expansion of the Universe, indicating classical instability here as well.

    \item Therefore, within a non-linearly interacting scenario, the EoS and SSS parameter asymptotes towards $-1$ from below without any finite-time divergences, indicating the little rip-like singularity.
\end{itemize}

\end{itemize}
In conclusion, for the linearly interacting case, $w_{\text{HH}}(t)$ diverges to large negative values for fractional $\alpha$ (e.g. $\alpha=1.1$ and $1.3$), while $v^{2}_{\text{HH}}$ stays negative throughout, signalling classical instability and a big‑rip–type finite‑time singularity rather than a little rip. In contrast, for the non‑linearly interacting case, EoS asymptotes to $-1$ from below for all $\alpha$, and SSS parameter remains negative but finite, giving a classically unstable yet little‑rip–like evolution without finite‑time divergences.


\bigskip

Let us now shift our attention to another late-time cosmological singularity, namely, pseudo-rip with the ansatz (\ref{pseudo rip ansatz}). For a pseudo rip, as the cosmic time tends to infinity $t\rightarrow\infty$, the $H(t)$ parameter approaches a finite value. For the ansatz established in Eq. (\ref{pseudo rip ansatz}), as $t\rightarrow\infty$ the Hubble parameter $H(t)\rightarrow H_{0}$ (positive finite value). As our next step, we will employ this ansatz in the EoS parameter (\ref{EoSHH}) and SSS parameter (\ref{vsHH}), and provide their plots in our subsequent discussions.

\begin{figure*}
    \centering
    \begin{subfigure}[b]{0.45\linewidth}
        \centering
        \includegraphics[width=\linewidth]{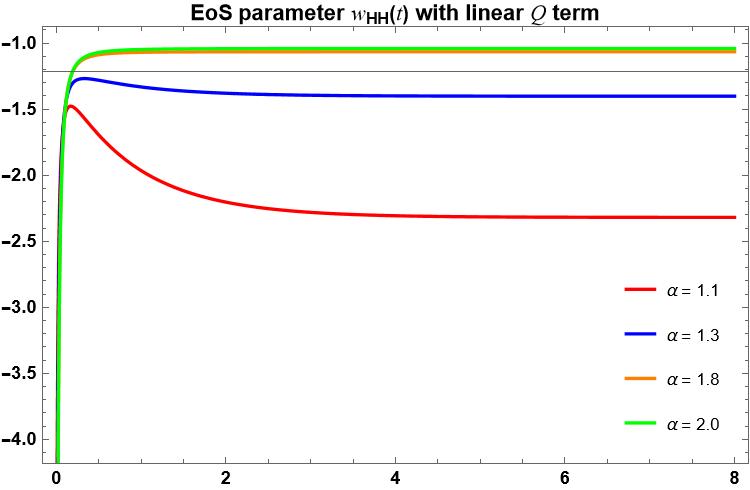}
        \caption{Plot for EoS parameter $w_{\text{HH}}(t)$ against cosmic time $t$ for various values of Lévy's index $\alpha$ within an ansatz established in Eq. (\ref{pseudo rip ansatz}) and $Q$ term in Eq. (\ref{linear Q}) for pseudo rip.}
        \label{Figure 5a}
    \end{subfigure}
    \hfill
    \begin{subfigure}[b]{0.45\linewidth}
        \centering
        \includegraphics[width=\linewidth]{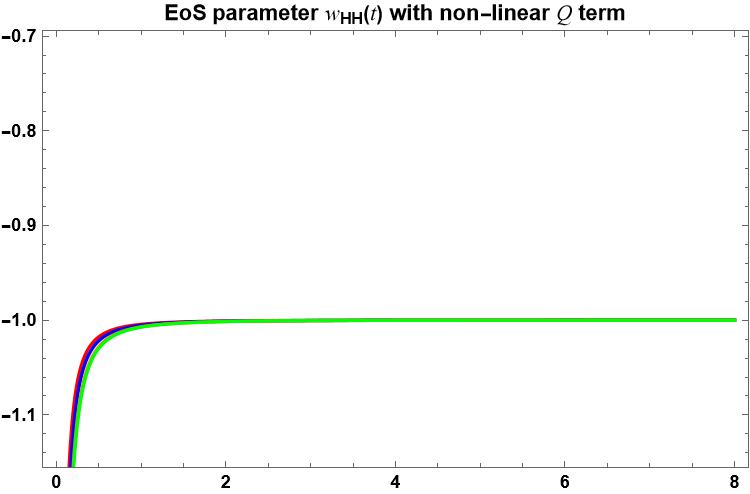}
        \caption{Plot for EoS parameter $w_{\text{HH}}(t)$ against cosmic time $t$ for various values of Lévy's index $\alpha$ within an ansatz established in Eq. (\ref{pseudo rip ansatz}) and $Q$ term in Eq. (\ref{nonlinear Q}) for pseudo rip.}
        \label{Figure 5b}
    \end{subfigure}
    \caption{Plots for EoS parameter $w_{\text{HH}}(t)$ for pseudo rip within Hubble cutoff.}
    \label{Figure 5}
\end{figure*}
\begin{figure*}
    \centering
    \begin{subfigure}[b]{0.45\linewidth}
        \centering
    \includegraphics[width=1\linewidth]{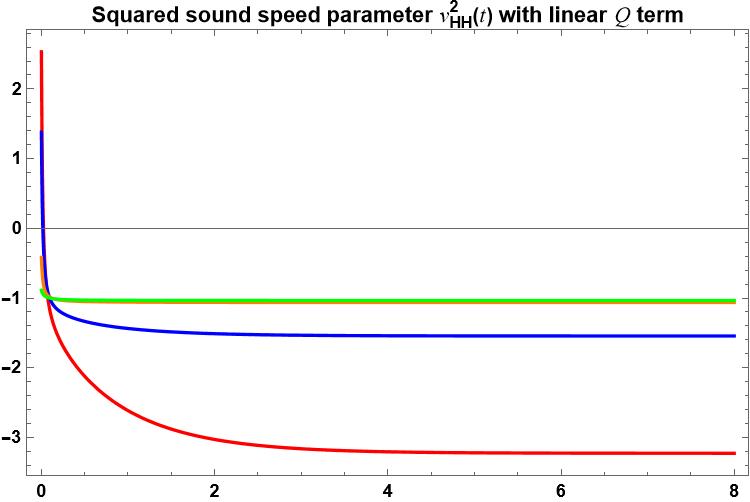}
    \caption{Plot for squared sound speed parameter $v^{2}_{\text{HH}}$ against cosmic time $t$ for various values of Lévy's index $\alpha$ within an ansatz established in Eq. (\ref{pseudo rip ansatz}) and $Q$ term in Eq. (\ref{linear Q}) for pseudo rip.}
    \label{Figure 6a}
    \end{subfigure}
    \hfill
    \begin{subfigure}[b]{0.45\linewidth}
        \centering
    \includegraphics[width=1\linewidth]{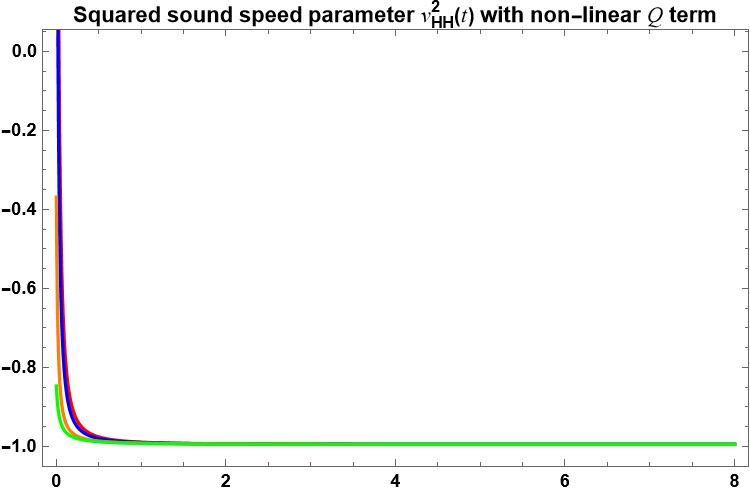}
    \caption{Plot for squared sound speed parameter $v^{2}_{\text{HH}}$ against cosmic time $t$ for various values of Lévy's index $\alpha$ within an ansatz established in Eq. (\ref{little rip ansatz}) and $Q$ term in Eq. (\ref{nonlinear Q}) for a pseudo rip.}
    \label{Figure 6b}
    \end{subfigure}
    \caption{Plots for squared sound speed parameter $v^{2}_{\text{HH}}$ for pseudo rip within Hubble cutoff.}
    \label{Figure 6}
\end{figure*}

\begin{itemize}
    \item Let us first elaborate on the linearly interacting scenario (\ref{nonlinear Q}) in Figures (\ref{Figure 5a}) and (\ref{Figure 6a}):

    \begin{itemize}
        \item In Figure (\ref{Figure 5a}), we observe that the EoS parameter asymptotically approaches a value closer and closer to $-1$ for large $\alpha$ values, while for smaller values of $\alpha$, the EoS parameter approaches a slightly lower value than $-1$. This behaviour for $\alpha$ does not represent pseudo rip.

        \item Within the requirements for a pseudo rip, the higher values of $\alpha$ provide a suitable pseudo rip-like behaviour as $w\rightarrow-1$ from below, making $\alpha\rightarrow2$ an appropriate parameter region.

        \item Furthermore, the corresponding SSS parameter interestingly starts from a positive value for smaller values of $\alpha$ but asymptotes towards negative values in the late-time limit, indicating classical instability.

        \item Therefore, the linear Q scenario plots show more dramatic evolution and large negative values of $v^{2}_{\text{HH}}$, which is less in line with the mild, asymptotic nature expected in a pseudo‑rip scenario.
        
    \end{itemize}

    \item Let us now elaborate on the non-linearly interacting scenario (\ref{nonlinear Q}) in Figures (\ref{Figure 5b}) and (\ref{Figure 6b}):

    \begin{itemize}
    
        \item From the evolution of the EoS parameter in Figure (\ref{Figure 5b}), one notices that for all values of $\alpha$, the EoS asymptotically approaches $-1$ from below, i.e., $w_{\text{HH}}\rightarrow1^{-}$ during late-time expansion.

        \item Since the EoS approaches $-1$ for all values of $\alpha$, one can say that a pseudo rip occurs for all values of $\alpha$.

        \item Furthermore, the corresponding SSS parameter interestingly asymptotes to $-1$ from above for all $\alpha$ in the late-time limit, indicating classical instability.
    \end{itemize}
\end{itemize}
In short, for the non-linearly interacting regime, the EoS parameter approaches $-1$ from below at late times, so the model generically produces pseudo-rip behaviour for all considered values of the Lévy index $\alpha$, as shown in Figure (\ref{Figure 5b}). For the squared sound speed, the linear-$Q$ case can cross from positive to large negative values (indicating classical instability), whereas the non-linear-$Q$ case asymptotes smoothly to $-1$ from above, giving a milder but still perturbatively unstable late-time regime.

\section{Discussions and Outlook}\label{Section 5}

The core idea of this paper was to investigate the interplay between Fractional Calculus (FC) and Holographic Dark Energy (HDE) framework during the late-time expansion of the Universe. Specifically, we investigated the role of non-local features introduced by FC in defining the occurrence or avoidance of rip singularities within the FHDE model. We constructed an approach to conduct such an investigation by making an explicit choice of a cutoff. Within our analysis, we make this choice to be the Granda-Oliveros (GO) cutoff and the Hubble cutoff. More specifically, we first investigated the occurrence of rip singularities for a GO cutoff, i.e., $L=(\gamma H^{2}+\delta\dot{H})^{-\frac{1}{2}}$, as it naturally incorporates terms like the time derivative of Hubble parameter $H(t)$, which enables a more efficient analysis of late-time behaviour than other cutoff prescriptions such as the Hubble cutoff, particle horizon cutoff and event horizon cutoff \cite{Trivedirip_2024, TrivediScherrer_2024}. On another informative note, we highlight that with the GO cutoff, rip scenarios such as big and pseudo rips can be easily studied without invoking any ansatz for the underlying Hubble parameter. In essence, the GO cutoff allows a broader perspective to investigate the existence or otherwise of rip singularities due to its evolved structure in comparison to other primitive cutoffs. In particular, the choice of a GO cutoff allowed us to explore the occurrence or avoidance of all the rip scenarios by analysing Eq. (\ref{Hint}). Based on our analysis, we find that a big rip \textit{does} occur for fractional values of the Lévy's index in the allowed range $1<\alpha\leq2$. We complemented this result with numerical illustrations of the big rip in Figures (\ref{Figure 1}) and (\ref{Figure 2}), which depict the late-time evolution of the EoS and Squared Sound Speed (SSS) parameters, respectively. Conversely, our results indicate that a pseudo-rip singularity could arise for $\alpha>2$, but we discard this case because Lévy's index is constrained not to exceed $2$. This clearly indicates that for a pseudo-rip singularity to be formed, the non-local features are not favoured, i.e., no fractional, while for the big rip singularity, non-local (fractional) features play a defining role. Furthermore, we emphasise that a little rip is expected to occur for a specific class of IR cutoffs for the FHDE model, and we underline that formation of a little rip singularity in this cutoff scheme is rather contrived and requires the cutoff to be $L\sim(\gamma H^{2}+g(H))^{-\frac{1}{2}}$, which belongs to the larger class of cutoffs, namely, the NO cutoff. With this understanding of rip scenarios in the GO cutoff, we proceed to section \ref{Section 4}, where we investigated the occurrence of little and pseudo-rip singularities through means of inserting the \textit{ansatz} for the Hubble parameter $H(t)$ that corresponds to the rip scenario at hand. For instance, see Eq. (\ref{little rip ansatz}) and (\ref{pseudo rip ansatz}) for the $H(t)$ ansatz of the little and pseudo rip singularity, respectively. In detail, we input the ansatz of $H(t)$ into the EoS, $w_{\text{HH}}(t)$, (\ref{EoSHH}) and SSS parameter, $v^{2}_{\text{HH}}(t)$, (\ref{vsHH}) derived for the Hubble cutoff, i.e., $L=H^{-1}$, within the FHDE framework, and numerically plotted their late-time cosmological evolution for a little rip in Figures (\ref{Figure 3}), (\ref{Figure 4}) and pseudo-rip in Figures (\ref{Figure 5}), (\ref{Figure 6}). Our primary reason for the choice of cutoff to be $L=H^{-1}$ was to demonstrate that even with the simplest form of cutoff, such as the Hubble cutoff, which has its own set of issues with conventional HDE, one can show the existence of late-time rip singularities, such as little and pseudo rips through means of fractional considerations. Our findings are summarised in Table \ref{Table}.

\begin{table}[h]
    \centering
    \caption{Table showcasing the occurrence (and avoidance) of rip scenarios with an ansatz-based approach.}
    \label{Table}
    \begin{tabular}{ccc}
        \hline
         Rip Scenario & Linear $Q$ & Non-linear $Q$\\
        \hline
        B.R (\ref{big rip ansatz}) & Occurs for $\alpha\rightarrow1.1$  & No B.R \\
        L.R (\ref{little rip ansatz}) & No L.R & Occurs $\forall\alpha\in(1,2]$ \\
        P.R (\ref{pseudo rip ansatz}) & No P.R & Occurs $\forall\alpha\in(1,2]$ \\
        \hline
    \end{tabular}
\end{table}

To summarise, let us first consider the EoS parameter for various rips:

\begin{enumerate}
    \item For the big rip scenario, we examine the late-time behavior of the EoS parameter $w_{\text{GO}}(t)$ within the GO cutoff (\ref{G-O}) in both linearly (Figure (\ref{Figure 1a})) and non-linearly (Figure (\ref{Figure 1b})) interacting regimes. In the linear $Q$ case, we find that $w_{\text{GO}}(t)$ evolves toward negative infinity, i.e., $w_{\text{GO}}(t)\rightarrow -\infty$ for $\alpha = 1.1$ and $1.3$ at late times, which signals the emergence of a big rip singularity in the linear interaction regime when fractional features start to dominate. Notably, the values $1.1$ and $1.3$ are illustrative, but in the true sense, for values $\alpha\rightarrow1.1$ provides a similar result, i.e., a big rip singularity, as shown in Figure (\ref{Figure 1a}). In contrast, the evolution of $w_{\text{GO}}(t)$ with a non-linear $Q$ term does \textit{not} support a big-rip-type singularity as the EoS parameter remains constant even at late-times. Hence, we infer the presence of a big rip singularity in the linearly interacting case when fractional features dominate in the limit $\alpha \rightarrow 1.1$.

    \item For the little rip scenario, we analyse the late-time evolution of the EoS parameter $w_{\text{HH}}(t)$ within the Hubble cutoff under linearly (Figure (\ref{Figure 3a})) and non-linearly (Figure (\ref{Figure 3b})) interacting regimes. In the linear $Q$ regime, we observe that the EoS evolution tends toward negative infinity as $\alpha\rightarrow1.1$, for values like $\alpha=1.1$ and $1.3$, which does \textit{not} signal towards the occurrence of a little rip. Whereas the evolution becomes constant as $\alpha\rightarrow2$ for values such as $\alpha=1.8$ and $2.0$, as shown in Figure (\ref{Figure 3a}). These values of $\alpha$ are illustrative, but a deeper insight reveals the \textit{absence} of a little rip singularity in the linear $Q$ case because the behaviour plotted in Figure (\ref{Figure 3a}) is reminiscent of a big rip singularity. On the other hand, the non-linear $Q$ case entails an evolution where for all values of $1<\alpha\leq2$ the EoS parameter asymptotically approaches $-1$ from below, which is a suitable evolution for realising a little rip singularity. Therefore, based on our analysis, we conclude that a little rip singularity occurs for allowed values of $\alpha$ in the range $1<\alpha\leq2$ within the non-linearly interacting regime.

    \item For the pseudo-rip scenario, we investigate the late-time evolution of the EoS parameter within the Hubble cutoff under linearly (Figure (\ref{Figure 5a})) and non-linearly (Figure (\ref{Figure 5b})) interacting regimes. In the linear $Q$ regime, we observe that the evolution of EoS parameter asymptotically approaches a value of $-1$ for $\alpha=1.8$ and $2.0$, while the EoS asymptotically approaches a value less than $-1$ for $\alpha=1.1$ and $1.3$, as shown in Figure (\ref{Figure 5a}). This directly implies that when fractional features are maximum, i.e., $\alpha\rightarrow1.1$, we do not have a suitable behaviour that corresponds to a pseudo rip singularity. Conversely, in the non-linear $Q$ regime, the cosmic evolution of the EoS is consistent with a pseudo-rip singularity, since for all values of $\alpha$ in the range $1<\alpha\leq2$, the EoS approaches $-1$ from below during late-time expansion of the Universe. Therefore, we conclude that a pseudo-rip singularity occurs for allowed values of $\alpha$ in the range $1<\alpha\leq2$ within the non-linearly interacting regime.
\end{enumerate}

Moreover, the squared sound speed (SSS) parameter remains negative for all rip scenarios within the FHDE framework during the late-time expansion of the Universe. This does not mean that the rip scenarios discussed here cannot occur; rather, it indicates that these FHDE-based rip scenarios exhibit classical instabilities at the perturbative level. Specifically, any perturbation in the fluid would induce superluminal propagation, signalling that the negative $v^{2}_{s}$ renders the system highly susceptible to even infinitesimal disturbances. This extreme sensitivity to initial conditions—characteristic of potentially chaotic dynamics—suggests that the long-term behaviour of rip scenarios in FHDE could be governed by non-linear instabilities, which may be quantified through Lyapunov exponents or similar measures of dynamical instability \cite{GOLDHIRSCH1987311}. To rigorously address these issues, various methods can be applied to investigate the evolution of perturbations in the FHDE framework (for a comprehensive review, see \cite{Astashenok_2024}). In particular, we emphasise that the analysis carried out by the authors in \cite{Astashenok_2024} can be generalised by incorporating fractional considerations and non-linear dynamical diagnostics in order to better understand the perturbation dynamics and stability thresholds within FHDE.

\bigskip

We highlight once more that this work adopts a purely classical perspective on rip scenarios in FHDE. A promising avenue for future research is a quantum cosmological treatment of the big rip singularity using the GO cutoff, which can be achieved by formulating a Wheeler–DeWitt (WdW) equation (see \cite{Albarran_2016, Bouhmadi-Lopez:2017ckh} and references therein for related discussions on this procedure).

\section*{Acknowledgements}

The authors would like to acknowledge networking support by the COST Action \textbf{CA23130} (Bridging high and low energies in search of quantum gravity (BridgeQG)). PM has also benefited from the activities of COST Action \textbf{CA23115} (Relativistic Quantum Information (RQI)). PM acknowledge the FCT grant \textbf{UID/212/2025} Centro de Matemática e Aplicações da Universidade da Beira Interior. OT would like to thank the Discovery Doctoral Fellowship at Vanderbilt for its support of the work.

\bibliography{main}

\end{document}